# Fostering Project Scheduling and Controlling Risk Management

**Abdul Razaque**
**Christian Bach**
**Nyembo salama**
**Aziz Alotaibi**

Department of Computer Science and Engineering
University of Bridgeport, USA.

**Abstract**

*Deployment of emerging technologies and rapid change in industries has created a lot of risk for initiating the new projects. Many techniques and suggestions have been introduced but still lack the gap from various prospective. This paper proposes a reliable project scheduling approach. The objectives of project scheduling approach are to focus on critical chain schedule and risk management. Several risks and reservations exist in projects. These critical reservations may not only foil the projects to be finished within time limit and budget, but also degrades the quality, and operational process. In the proposed approach, the potential risks of project are critically analyzed. To overcome these potential risks, fuzzy failure mode and effect analysis (FMEA) is introduced. In addition, several affects of each risk against each activity are evaluated. We use Monte Carlo simulation that helps to calculate the total time of project. Our approach helps to control risk mitigation that is determined using event tree analysis and fault tree analysis. We also implement distribute critical chain schedule for reliable scheduling that makes the project to be implemented within defined plan and schedule. Finally, adaptive procedure with density (APD) is deployed to get reasonable feeding buffer time and project buffer time.*

**Keywords:** Fuzzy failure mode and effect analysis (FMEA), Risk management, Monte Carlo simulation, critical chain schedule, schedule reliability, time buffer.

## 1. Introduction

The manager is responsible for a project to focus for answering the fundamental questions related to the project throughout the different phases of project. There are two main issues that can be defined as project duration and an uncertainty subject of task duration. In reality based on project, the critical approach is always changed critic of with progress of project, and the project manager has no prediction to project future time. Therefore, risk management is very necessary to project success and it is as a critical component of managing any project [1].
Risks are very important issues that may influence to achieve objectives of project within required time. Risks are not under control of project team because they are depending on future events. The major risks are mitigated and planned during the project life cycle. The organizations face the different types of risks include stakeholder struggle constitutional changes, difficulty budget reduction, project plan cost, project duration, service provider reliance and process change requirements [3].

The significant way of controlling risks in projects is to build up reliable project estimates and schedules [2]. The scheduling optimization is viewed from a reliability perspective. The risk management allows checking the project risk as a constraint. It is assumed that a liaison always exists between existing risks resided in a project and related activity's time. To accomplish the goal of this paper, the combination of risk management and critical chain schedule model is applied to choose the best schedule reliability for a project.

## 2. Managing Risk Schedule

Since risk management is the analysis, assessment, control of expositing to the risk in order to reduce such disaster. To minimize the risk impact and to manage responses is to do materialize (contingency plan). It is in a need to provide funds to cover risk impact that might already materialize [3]. Identification, analysis, evaluation, response, and monitoring are the principles of recognizing the risk management process.

There are three factors that should be taken in account to avoid the management risk: a tight time constraint, uncertainty and structural complexity. Exceeding the project time is one of the most important effects in the project period. Having a Schedule risk management that can help reduce the risk is the main factor of success during the implementation of the project. In addition, using efficient risk analysis method helps the risk management to obtain an accurate result that would help in making the right decision to avoid the risk.

## 3. *Analysis of Risk*

The purpose of analyzing risk is to help reduce the risk, and also to anticipate what would happen in the future if any decision might be taken in organization [8]. There are some techniques to achieve risk analysis of projects: first, qualitative technique such as Analytic Hierarchy Process (AHP) and risk matrix. Second, quantitative technique such as Fuzzy Logic , Tree, Monte Carlo Simulation, Failure Mode and Effect Analysis (FMEA), Event Tree Analysis (ETA), Fault Tree Analysis (FTA). To calculate probability and severity and getting an accurate result, quantitative analysis should be used to have an effective risk assessment.

The advantages and disadvantages of these techniques are explored to discover efficient and applicable methodology. However, quantitative analysis cannot be used universally according to their limitations. The Risk Priority Number (RPN) has no ability to distinguish between the importance of the input variable such as Severity, detection, and occurrence while calculating the RPN. Fuzzy FMEA and FTA and ETA are a complete and efficient method that helps to find the main reasons behind causes (JRAP) judgmental risk analysis process is a new schedule risk analysis method that has been proposed and implemented through different project using the duration equation JRAP. JRAP is a pessimistic risk analysis methodology that is effective in scheduling uncertain conditions for the risk management. The focus of this method is to evaluate the critical risk impact through the duration of the e project. In addition, risk management can help the project to be completed on time and also to provide schedule reliability. Critical chain schedule is one of the efficient that can help out projects to be accomplished earlier and with better reliable scheduling.

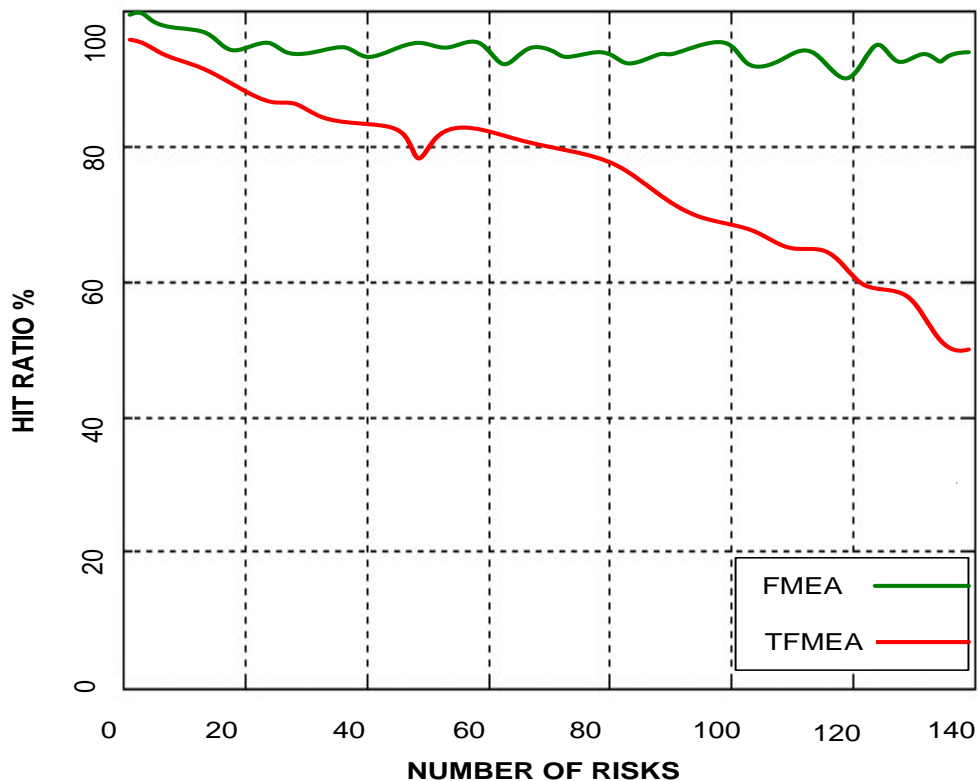

**Figure 1: Comparison between traditional FMEA (TFMEA and FUZZY FMEA**

## 4. Critical Chain Estimation

Projects consists of some tasks that should be done on time and these tasks control the estimate of the duration, furthermore, the task planner has to considered the safety factor to ensure the completion of the project on time[14]. This factor might add overestimated time that means the project should be completed before what has been scheduled: this added time is not needed, and sometime is wasted time. The Critical Chain approach helps to gain more predictability, productivity, reliability and speed in order to have a better outcome [15].

## 5. Analysis of Time Buffer Approaches

To determine the best buffer size method is to know that the buffer time might be needed to complete the project on time. Statistical fluctuations, project characteristics, buffer management, and managements' experience give the project manger a better vision over the project. C&PM (Cut and Paste Method), RSEM (root square error method) and adaptive buffer sizing methods are the buffer sizing methods that are used to generate buffer size that would help improving the chances of completing on time. The aggregation of the risk that has been encountered along the tasks is a kind of buffers sizing that is not a scientific method such as The Cut and Paste Method (C&PM) and the RSEM methods. Buffering size has been suggested. Figure 2 shows on intuitive assessment of risk.

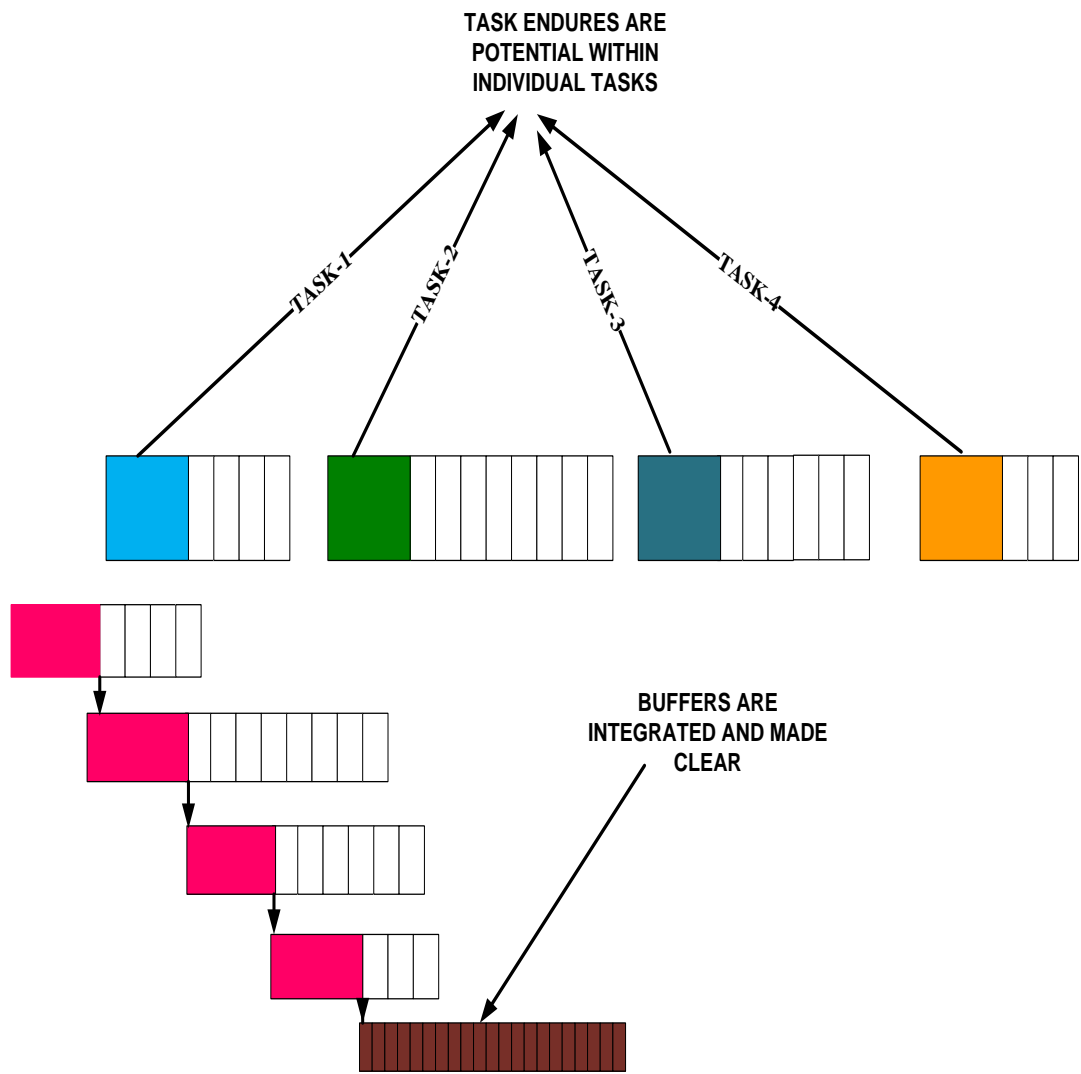

**Figure 2: Showing buffer sizing and its use.**

**A. Cut and Paste Method (C&Pm)**

In case, safe factor that has been estimated for each task is already assumed and given to calculate the critical chain and feeding chains, the task duration would have safe estimates.

After determining the critical chain, summing the whole safety cut from the feeding chain and uses it a feeding buffer as a half of the sum. Critical chain merges with feeding chain which is the feeding buffers that has been added to the end of the feeding chain [19].

**B. Root Square Error Method (Rsem)**

Two estimates are used to determine the critical chain in this method like C&PM, average estimate and the safe estimate [17]. We calculate duration for uncertainty with help of equation (1).

$U_k = S_k - A_k$     (1)

Where,

$U_i$ is uncertainty of task k, $S_k$ is safe estimate of task k and $A_i$ is average (60%) estimate of task ki, for all k in chain of feeding [17]. It suggests that standard deviation can be determined in task duration with ($U_k/2$). Hence, feeding chain calculate standard deviation by equation (2).

$$\beta = \frac{\sqrt{(\frac{U1}{2})^2 + (\frac{U2}{2})^2 + \ldots, (\frac{Uk}{2})^2}}{2a}$$     (2)

Where,

'K' is feeding chain activities. Assume task completion times are to be independent.

Buffer size and two standard deviations are shown in (3).

Buffer-Size $= 2\beta = \sqrt{(U1)^2 + (U2)^2 + \ldots, (Uk)^2}$     (3)

The sizing methods has been discussed in [19] and obtained estimated 60% of the variability in work for the equivalent period. However, as [14] discussed the 60% variability in the distribution of job durations is skewed to the right generally does not reflect important. Thus, in determining the buffer size to durations should be clear about the basic assumptions. Moreover, the total resource usage is close to the total resources available, will likely delay that has occurred. Thus, there should be a delay buffers to absorb. Similarly, for a given number of tasks, precedence relations, as volume increases, it is again likely that delays will occur. Most activities are interrelated and work in the case of a delay in completion will affect all of its successors. Therefore, the buffer size increases the number of precedence relations should increase.

**C. Adaptive Procedure With Denisty Methos Method (Apd)**

Walter, Rom and Sandra [19] proposed the Adaptive Procedure with density (APD) to address C&PM and RSEM method limitation. In this method, total number of precedence relationships (Tpr) on the sub-network feeding into the critical chain, NUMTASK is total number of tasks on the sub-network and the variance of activity ($VA_i$). Then equation 4 illustrate, for each feeding chain:

$FC = 1 + \frac{Tpr}{Tn}$     (4)

Where
SUM = 0

For every activity i on the longest path terminating at the critical chain calculated through equation 5:

SUM = SUM + $VA_i$

Size of buffer = FC * $\sqrt{SUM}$     (5)

The value of $VA_i$ depends on the assumption regarding the distribution of task durations. The APD method has been tested with Patterson data set which is a full factorial experiment data [20]. The simulation results indicate that C&PM planned completion times will be 17–25% longer than the adaptive method. This might cause a 4-year long project to have a planned completion time of 5-years plus a possible project buffer. The RSEM generates results between these two extremes. A project manager who plans to use a method like RSEM should instead also include project characteristics as well and use something more sophisticated such as APD [19].

Thereby, the project manager main objective when selecting buffer sizing method should employ the one which generates a schedule with shorter project completion time but which can be met with a high probability.

## *6. Research Methodology*

The primary goal of risk management is to focus on the actual risk that can be essentially identified or analyzed as a risk in the project but it is absolute mitigating in all form of the project risk that often occur [23]. Therefore, risk analysis point out to be involved in the multiple risks with low or high degree of RPN. It is logically an effect on project time that has not been completely considered in the form of risk analysis. To be able of solving this kind of the issue, risk management should be immediately incorporated with other methods to amplify schedule reliability. Critical chain schedule (CCS) is defined as first departure from empirical project management. By selecting critical path in CCS, it does have a few strong capability to identify, and analysis risks, so that risk management can be freely and useful to detect risky activity in project plan [22]. Hence both methods together help project to be ended up in the time manner which is most useful factor to project success. In this research paper, the proposed method consist of a number of managerial step to be executed and an equation modeling that explain all variation in each activity of duration in the scheduled network. This methodology is first defined as an integration of risk management and it is secondly described as a critical chain schedule in order to promote project completion in time.

### A. Fuzzy Fmea and Fuzzy Ahp

The first task undertaken during the conduct of this methodology is to find out the critical risks. To identify and analyze the risk, Integration of fuzzy logic and FMEA is proposed to define the probability of occurrence (P), impact (I), and detection/control (D). Each variable is defined using membership functions (MFs) over the universe of discourse of 1 to 10 and five linguistic terms (Very High, High, Low, Very Low) for value of them. Figure 3 shows the membership functions of probability of occurrence (P) as an example

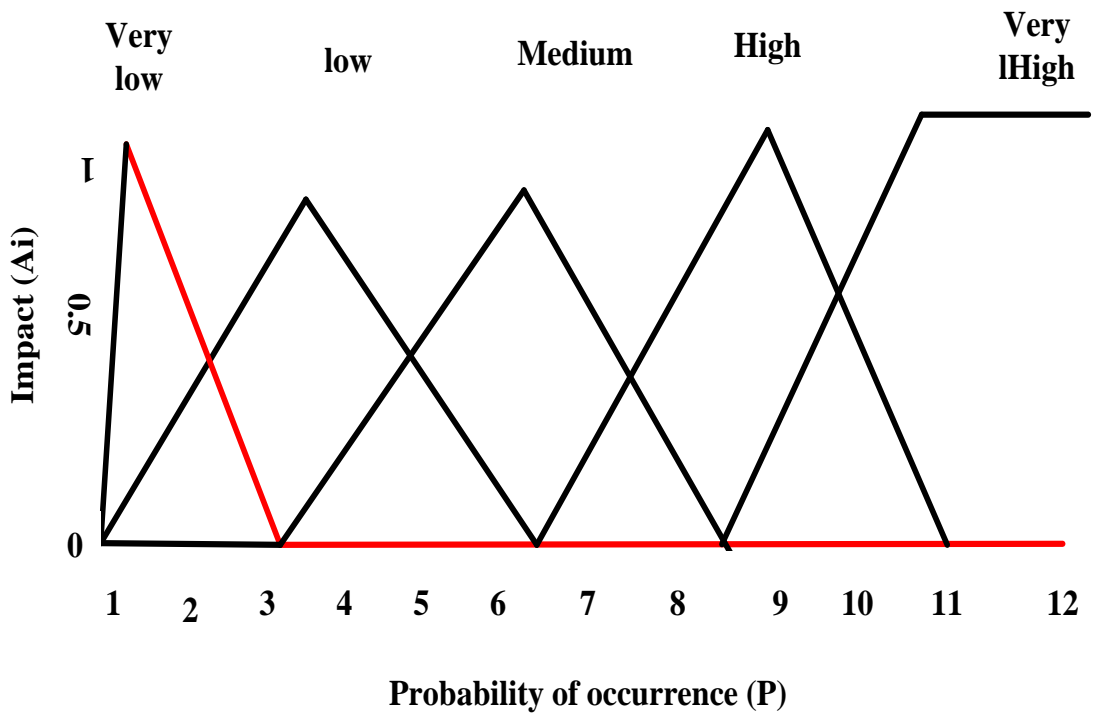

**Figure 3: Membership functions for probability of occurrence (P)**

In order to calculate impact of each risk, three factors should be considered. Impact of cost, time and quality. For this purpose the fuzzy AHP approach has been adopted to solve the multi-criteria decision-making problem by integrating of three factors into one variable named aggregated impact (AI), which is calculated in equation 6.

AI=TPC * IC+TPT*TI+TPQ*IQ          (6)

Where,

TPC=Total priority cost, IC= Impact of cost, TPT= Total priority time, TI=Impact of time, TPQ=Total priority quality and IQ= Impact of quality.

Total priority (TP) of each criterion is calculated by (7), Now take an average of summation of $X_{ij}$.

$$AI = TPC * IC + TPT*TI + TPQ*QI \qquad (7)$$

$TP_i = \sum X_{ij}/3$ , i = j = quality, cost and time.

$X_{kl}$ is relative factoring importance k over l that is calculated by equation 8:

$$X \frac{l+2*(m+n)+o}{6} \qquad (8)$$

B and C represents a minimum value, represents the highest potential, and D represents the maximum. The possibility of the presence (P), impact (I), and detection / control (D) assessment, risk significant number (RCN) should be calculated. Membership functions for the RCN to choose. It is important to note that any risk assessment is an event that RCN [11] falls within the definition of risk.

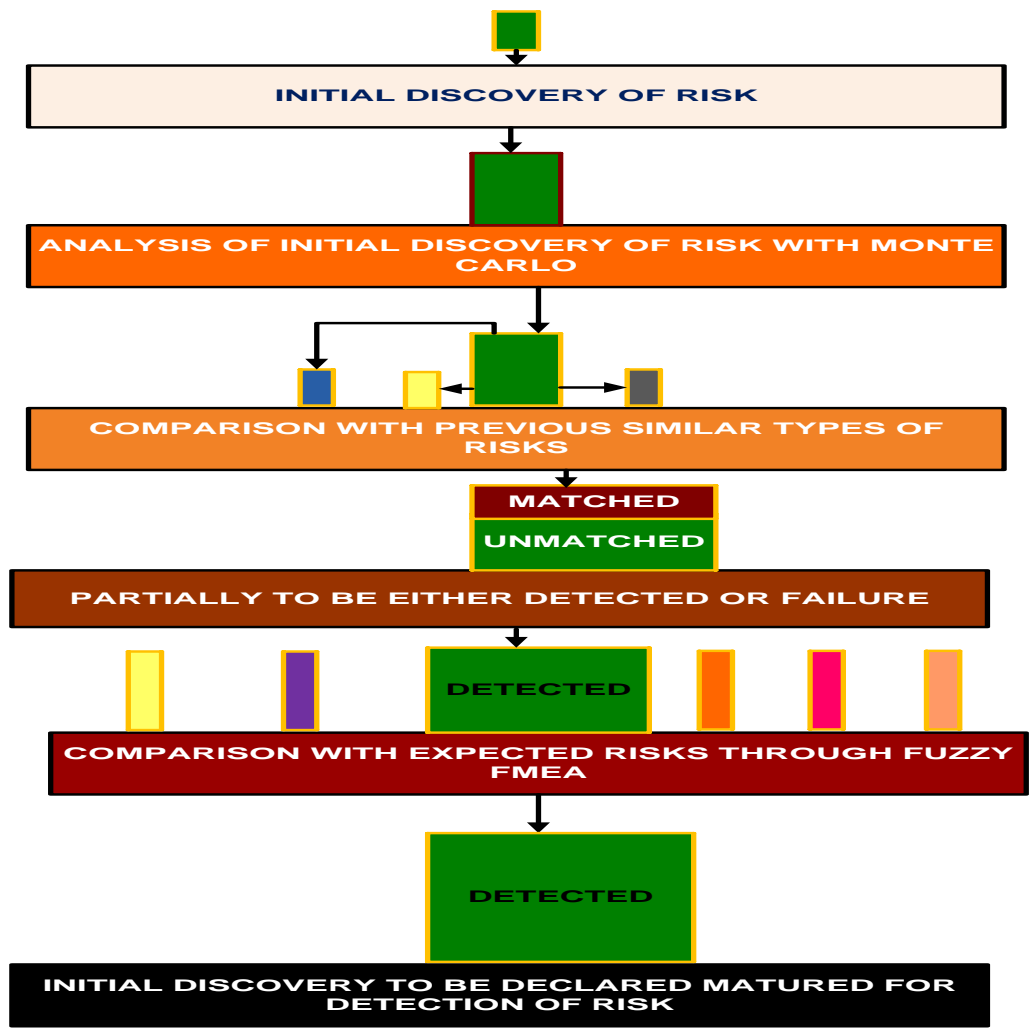

**Figure 4: showing correct method for detection of risk**

The purpose of using RCA is to support the decision-makers in assessing the level of risk criticality. It reads P, CI, TI, SI, and D from the risk register for each risk event, calculates the aggregated impact (AI) using fuzzy AHP, exports P, AI, and D to the fuzzy expert system to calculate the RCN, and presents the resultant RCN. Figure 4 shows how to detect risk in any project.

**B. Modelling and Simulation**

At this step, Monte Carlo simulation (MCS) utilizes the activity-risk factor matrix developed at the third step in order to calculate the variation in activity durations.

In order to enable MCS to calculate the activity durations in a stochastic manner, the following pessimistic is presented in equation 9. Where, Min. Time is the minimum duration that an activity can be completed, Max. Time, the maximum duration that an activity can be completed $RF_n$, the percent effect of nth risk factor on an activity (taken from activity-risk factor matrix), Random is a random number, between 0 and 1, generated during MCS to represent the nth risk factor's probability distribution. Once this process ends, it is possible to calculate total project time and also critical path [24]. A fuzzy model utilizes fuzzy rules, which are linguistic statements pertaining to fuzzy sets, fuzzy logic and fuzzy inference. Fuzzy rules play a critical role in representing expert control, modeling knowledge and experience in linking the input variables of fuzzy controllers to output variables. Two major types of rules,viz. Mamdani fuzzy rules and Takagi-Sugeno are available for modeling a fuzzy rule base.

Utilizing such models that would identify actual defects and relying on fuzzy logic techniques in order to generate test patterns and create fault simulation we can limit risks that would negatively impact project completion in the future. Reliance on integration of Fuzzy FMEA and Fuzzy AHP creates realistic fault coverage in comparison to the conventional methods. The simulation results include the fault coverage and test-pattern statistics which would be beneficial in limiting risks. A model is automatically similar to simpler than the major system it represents. Model enables the analyst to predict the effect of changes to the system. On the one hand, a model should be a close approximation to the real system. On the other hand, it should not be so complex than it is impossible to understand and experiment with it [16]. The best model is a cautious tradeoff between two words simplicity and realism. But, Simulation of a system is characterized by the operation of a model of the system. Simulation is the simple tool to evaluate the performance of a system, existing or proposed, under different [21].
 It is often used before an existing system is altered or a new system built, to reduce the failure to meet specifications, to eliminate unforeseen bottlenecks, to prevent under or over-utilization of resources, and to optimize system performance.[16] & [21].

Time of activity = $MAX[(Min\_time + Max\_time - Min\_time) * (FR_{1+} R_1) + (FR_2 * R_2) + \ldots + (FR_n * R_n)]$, $[Max\_time - (Max\_time - Min\_time) * (FR_1 * R_1) + (FR_2 * R_2) + (FR_n * R_n)]$                    (9)

**C. Risk Mitigation**

In this step some actions to prevent from delaying in project and protect a system from failure has been considered. Towards this, the utilization of the fault tree structure allows experts to understand the root causes of the risk event and supporting experts to understand which root causes are contributing the most to the occurrence of the risk event and what the fuzzy probability of the CRE is. Besides, the mitigation strategies are established that can either eliminate or reduce the chances of the occurrence of the highest ranked root causes. Event tree analysis offers for risk mitigation and understanding the impact on a risk event by considering the failure and success of the identified mitigation strategies. the fuzzy probability of failure of each mitigation strategy has been done with FTA and The probabilities of success branches are evaluated as (1-the probability of failure).eventually determine the overall probability of each path by multiplying the fuzzy probability of all the events located on the same path. This method helps improve safety integrity and minimize the risk. Figure.5 illustrated the combination of FTA and ETA.

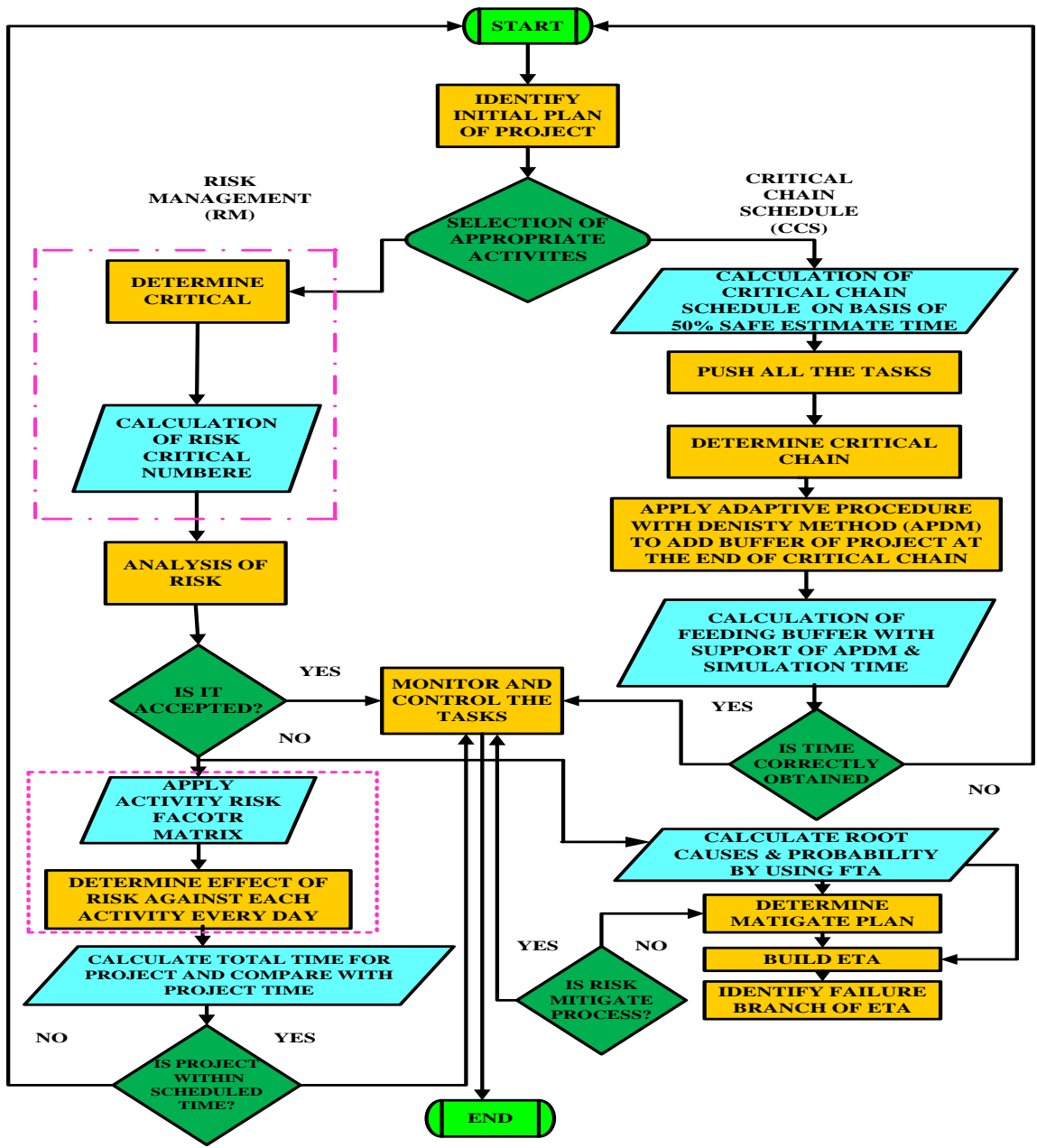

**Figure 5: Monte Carlo Simulation and Fuzzy FMEA**

Risk Mitigation can be represented as the second step of the risk management process. It plays the role of involving, prioritizing, evaluating, implementing and maintaining the suitable risk reducing controls or measures that it is mainly recommended from the risk assessment process or risk Assessment Compliance Component [26]. Risk Mitigation can also be enabling management to reduce mission risk, when it is feasible. It influences managers to change the operations and the costs of protective measures. In addition, it achieves gains in mission capability to protect the agency systems, and information that support their missions. In addition, missed dependencies within schedules can be a serious risk to the project which makes the identification of possible risks in the path of delivering successful project a necessity. Risks to the successful completion of the project are managed through the "backward" building of the network which understands the required outputs prior to defining inputs. Although the project buffer and its security from critical chain variability have undeniable importance, feeding buffers require equal attention. Buffers work towards securing project promises from risks involving parallel activity.

Projects are primarily the products of several integrations and the changing of critical paths and project delays are a common phenomenon. In traditionally managed projects, integration risk, which can be defined as the statistical nature of merging parallel paths, is the primary source of changing critical paths. The Monte Carlo simulation is a commonly used tool in risk assessment and thereby mitigation as it provides the potential impact of integrations on the project timeline. Variability in feeding chains is dealt with by the building of feeding buffers in the he critical chain schedule. As Monte Carlo simulations predict the probability of achieving project goals and meeting project deadlines, buffered critical chain schedules are customized to reduce integration risk and act as a risk mitigation measure.

**D. Critical Chain Schedule and Improving Consistency**

The Critical Chain Schedule ensures the efficient transition of a dependency network into a relatively low risk project. Resource as well as handoff dependencies are systematically included in the determination of the critical chain which in turn allows for greater reliability. Additionally, estimates developed during the planning stage are utilized to redirect safety measures to secure the project's goals and potential value [25].

## *Conclusion*

The goal of this paper is to introduce risk management and critical chain methodology that is pertinent for all projects. The conceptual mathematical model for assimilation of risk management has been proposed with critical chain schedule. The methodology used for project consistency of scheduling resolves many risk management issues. Risk management process categorizes analysis and mitigation of the risk. There is no contemplation of risk impact on project that does not provide solid actions for improving the reliability of project scheduling. Since, leading methods like critical chain schedule and theory of constraint and critical chain schedule have no obvious methodical technique to determine and analyze the risk. Therefore, the proposed model in paper can be emerged to locate boundaries of constraints with proficient and realistic methodology to schedule risk management in engineering and industrial projects.